\documentclass{elsart}
\journal{New Astronomy}
\usepackage{natbib}

\usepackage{graphicx}
\usepackage{epsfig}

\usepackage{amssymb}

\begin{document}

\begin{frontmatter}

% Title, authors and addresses

% use the thanksref command within \title, \author or \address for footnotes;
% use the corauthref command within \author for corresponding author footnotes;
% use the ead command for the email address,
% and the form \ead[url] for the home page:
% \title{Title\thanksref{label1}}
% \thanks[label1]{}
% \author{Name\corauthref{cor1}\thanksref{label2}}
% \ead{email address}
% \ead[url]{home page}
% \thanks[label2]{}
% \corauth[cor1]{}
% \address{Address\thanksref{label3}}
% \thanks[label3]{}

%\title{Clover}
 \title{Clover - A B-mode polarization experiment}

% use optional labels to link authors explicitly to addresses:
% \author[label1,label2]{}
% \address[label1]{}
% \address[label2]{}

\author{Angela C. Taylor$^1$ for the Clover collaboration}
\address{$^1$Oxford Astrophysics, Department of Physics, Denys
Wilkinson Building, Keble Road, Oxford, OX1 3RH, UK}

\begin{abstract}
% Text of abstract

Clover is a new instrument being built to detect the B-mode polarization of
the CMB.  It consists of three telescopes operating at 97, 150, and 220 GHz
and will be sited in Chile at the Llano de Chajnantor.  Each telescope
assembly is scaled to give a constant beam size of 8 arcmin and feeds an
array of between 320 and 512 finline-coupled TES bolometers.  Here we
describe the design, current status and scientific prospects of the
instrument.
\end{abstract}


\end{frontmatter}

% main text
\section{Introduction}

The polarization of the cosmic microwave background (CMB) contains
unique information about the early universe. In particular, the
presence of a tensor (gravitational wave) component, which can only be
determined with limited accuracy from the temperature anisotropies,
produces a unique signature in the CMB polarization, the B-mode
signal. The polarization field can be decomposed into curl-free
(E-mode) and curl (B-mode) components, and the polarization from
scalar fluctuations produces no B-mode in linear theory. Provided
spurious sources of B-mode signal (such as foregrounds and systematic
effects) can be controlled, detection of a B-mode signal would thus
constitute a direct measurement of primordial gravitational waves and
would measure fundamental parameters of inflation theory. However, the
expected amplitude of the signal is exceedingly small ($< 0.1\mu$K)
and very precise control of systematic effects, as well as raw
sensitivity, will be needed to measure it reliably.

Clover is a project to attempt to measure the B-mode polarization of
the CMB. It is a collaboration between the Universities of Oxford,
Cardiff, Cambridge and Manchester, with the detector read-out
technology supplied by the University of British Columbia and the
National Institute of Standards and Technology. Clover consists of
three separate, scaled instruments operating at 97, 150 and 220 GHz to
provide discrimination between the CMB and foregrounds. The telescopes
use the Compact Range Antenna design, which has particularly low
aberrations across a large focal plane. The telescope focal planes
will be populated with arrays of polarimeters, in which the
polarizations in each pixel are rotated, separated and detected so as
to extract the Stokes parameters Q and U, modulated by the
polarization rotation. Intensity can also be recovered, but only
modulated by the telescope scanning across the sky. The signals will
be detected using arrays of superconducting transition edge sensors
(TESs), which allow background-limited sensitivity and can be
fabricated in sufficient quantity using lithographic techniques.

It is planned to site Clover at Chajnantor in Chile, where the atmospheric
transparency is excellent, and the latitude allows for
constant-elevation scans of a single sky area to be cross-linked. The
beamwidth at each frequency will be 8 arcmin and the telescopes will
be scanned to give sensitivity over an angular multipole range of $20 <
l < 1000$. This will allow measurement of the B-mode signal over the
range of scales expected for the signal from primordial gravitational
waves. The sensitivity should be around 220$\mu {\rm K} s^{1/2}$ per pixel at 150
GHz, sufficient for the sensitivity to be limited by the B-mode
lensing foreground after two seasons of observation. This will result in
a sensitivity to gravitational waves equivalent to a
tensor-to-scalar ratio, $r$, of $r > 0.01$.

\section{Instrument design}
\label{sec:inst}

The principles behind the Clover design are to have the cleanest
possible optical system, a large number of highly sensitive focal
plane detectors, and multiple levels of modulation of the signal to
differentiate between astronomical and systematic signals.

\subsection{Telescope Optics}

Clover uses the Compact Range Antenna design (also known as a crossed
Dragone) which is an off-axis un-blocked aperture design using a
concave hyperboloidal secondary and a paraboloidal primary. Despite
being off-axis this system has a very large focal plane with low
aberrations and cross-polarization \cite{Yassin-CRA}, much more so than
the equivalent aperture Gregorian, for example. The drawback is that
the secondary has to be roughly the same size as the primary, but this is not
a serious issue for a modestly-sized ground-based telescope.

\begin{figure}[h]
\begin{center}
\includegraphics[height=0.3\textheight,angle=0]{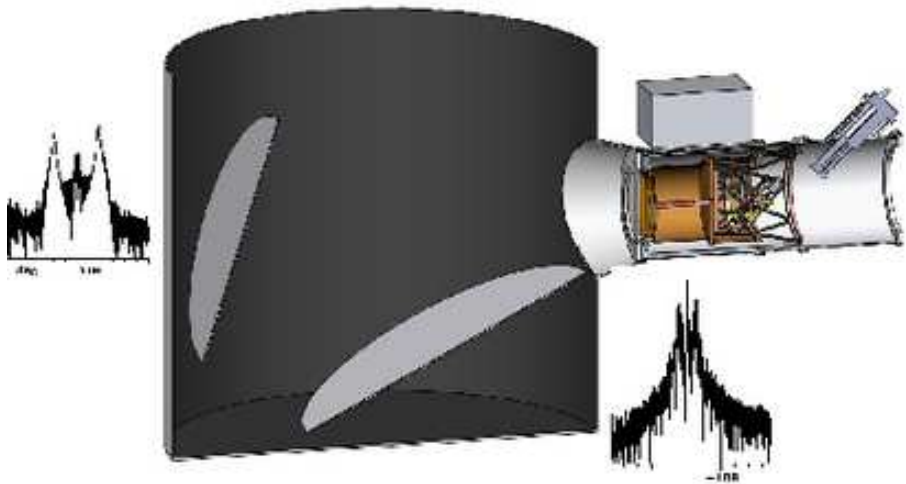}

\end{center}
\caption{\label{shield}
Artists impressions showing how an absorbing inner groundscreeen will intercept the residual far-out sidelobes. }
\end{figure}

The Clover 97-GHz system uses a primary with a 1.8-m projected
aperture. This is significantly under-illuminated to minimise
sidelobes and spillover, giving a beam of about 8 arcmin. The far
sidelobe levels of the basic optical system are extremely low over the
whole sphere (typically -80 dB), apart from spillover around the two
mirrors (which is less than 1 percent of the total power). This
spillover signal is absorbed by a cylindrical absorbing screen
surrounding the optics (see Fig. \ref{shield}). This screen adds 5--10
percent to the power load on the detectors, but prevents polarized
sidelobes from reaching the ground or bright astronomical sources.

Each optical assembly (one for each frequency) is mounted on a
three-axis mount (azimuth, elevation and boresight rotation) which
will be covered with a retractable dome to provide protection when not
observing. The mount provides full sky coverage down to the horizon,
azimuth slew speeds up to 20 deg/s, and over 180 degrees of boresight
rotation (although in practice the useful degree of boresight rotation
is limited by the need to keep the pulse-tube cooler of the cryostat
within $\sim 45$ deg of vertical).

\subsection{Cryogenics}

The Clover detectors need to be cooled to less than 100 mK in order to
provide optimum performance. This is achieved using three cooling
mechanisms: a Cryomech PT-410 pulse-tube cooler which provides 40-K
and 4-K stages, a high-capacity Simon Chase He-10 cooler providing
2-K, 370-mK and 250-mK stages, and a custom-made miniature dilution
refrigerator \cite{Tele06}, providing final cooling of the detectors to
60 mK. The pulse tube cooler has the motor and expansion stages
separated by flexible stainless steel lines to minimise vibrations in
the cryostat (although TES detectors are much less sensitive to
microphonics than, for example, NTD bolometers). The temperature of
the cold stages will be controlled using a closed-loop system with a
long-term design stability of 3.5~$\mu$K rms. A stack of metal-mesh
filters provides bandpass filtering and infra-red blocking: the size
of the cryostat window is limited by the largest such filters that can
be manufactured, to 300~mm. This fixes the largest number of feedhorns
that can be used; at 97 GHz this is 160, but at the higher
frequencies, more than the planned number of 256 horns could be
accommodated.

\subsection{Feedhorn Arrays}

The feedhorns are corrugated, profiled horns, which have a steeper
beam fall-off than conical corrugated horns, and hence give less
spillover, particularly for the edge feeds in the array. Each horn is
approximately 18~mm diameter and 100~mm long (at 97 GHz; scaled at the
other frequencies). The horns will be individually electroformed and
bolted to the polarimeter block.

\subsection{Polarimeter}

The polarimeter has to provide for separation of the orthogonal
polarizations and their modulation. We will describe here the 97-GHz
system, which is the most advanced in design. Following the feedhorn,
the signal passes through a mechanical waveguide polarization rotator,
consisting of a `half wave plate' section, implemented with iris
polarizers in circular waveguide, which has extremely low insertion
loss and return loss, and very flat phase response across the whole
(30 percent) band. Following the polarization rotator, the
polarizations are separated in a waveguide orthomode transducer, which
outputs two rectangular waveguides. These interface to the detector
block, in which each polarization is detected with a separate
finline-coupled TES chip.

At the higher frequencies it is more difficult to implement the
mechanical rotating waveguides, and we are considering several
options, including a single rotating waveplate in front of the horn
array as well as planar circuit phase switches \cite{phase-switch}. In
this latter design, a switch is implemented using a superconducting
nanostrip shorting a transmission line; the device can be made to
switch between reflection and transmission by biassing the nanostrip
with a current exceeding the superconducting critical current
density. Given such a switch, there are several ways of implementing a
transmission-line phase switch which can insert a 90- or 180-degree
phase shift. The polarimeter would then consist of a phase switch in
each of the OMT outputs, followed by a hybrid coupler and then the
detectors. Although more complicated in principle, such a polarimeter
could be implemented entirely on a single lithographed chip.

\def\plotfiddle#1#2#3#4#5#6#7{\centering \leavevmode
\vbox to#2{\rule{0pt}{#2}}
    \includegraphics{#1}}

\subsection{Detectors}
\label{sec:detectors}
Clover uses finline-coupled superconducting transition-edge sensor (TES)
bolometers as detectors.  To detect the two polarizations the 97-GHz
telescope has 320 detectors while the 150 and 220-GHz telescopes have
512 detectors each. The detectors are cooled to 100~mK in order to
achieve the target NEPs (1.5, 2.5, and $4.5\times10^{-17}\rm\ W/\sqrt
Hz$). Each detector is fabricated as a single chip to ensure a 100\%\
operational focal plane.  The detectors are contained in linear
modules made of copper which form split-block waveguides and are read
out with time-division SQUID multiplexers \cite{Cher99,deKorte03}
fabricated at the National Institute of Standards and Technology.
Further amplification of the multiplexed signals is provided by SQUID
series arrays\cite{Welt93}.

\begin{figure}[!b]
\vskip.5cm
\includegraphics[width=6.0cm]{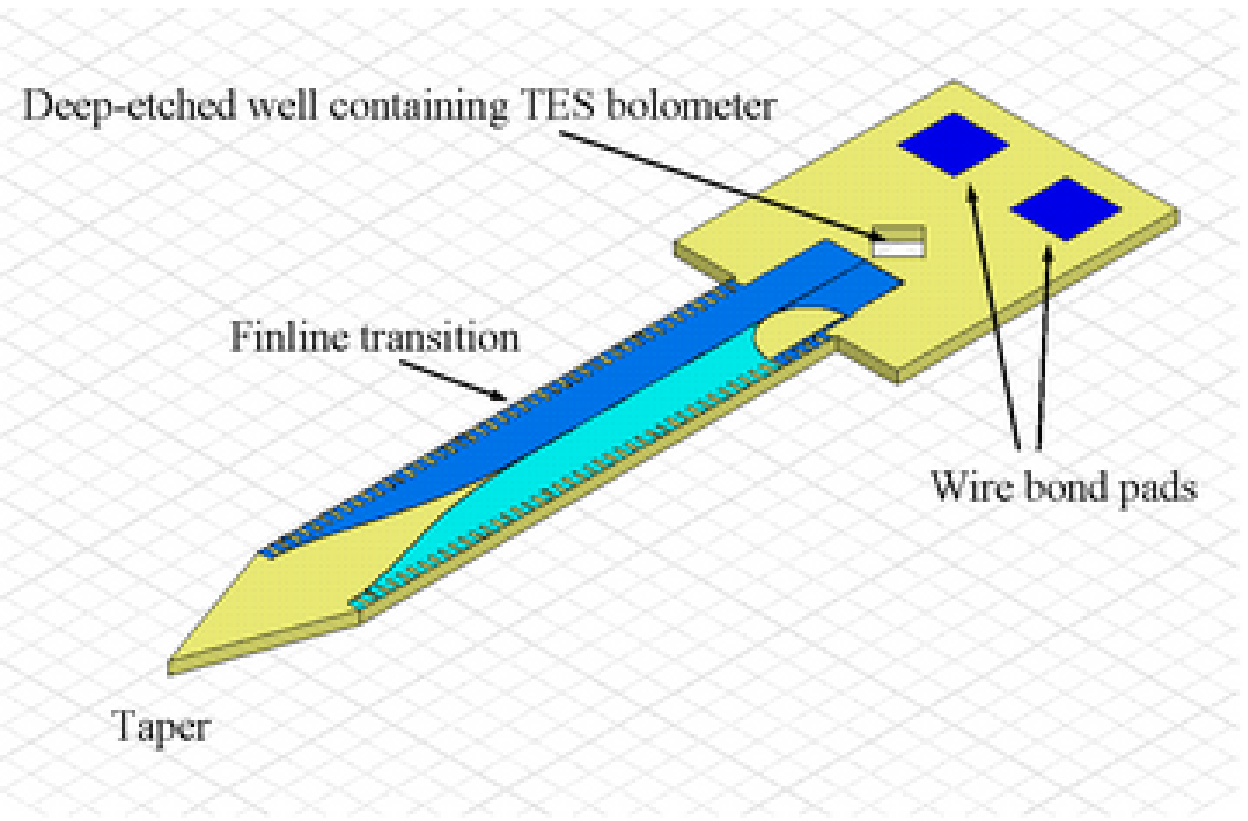}
\hspace{1cm}%
\includegraphics[width=6.0cm]{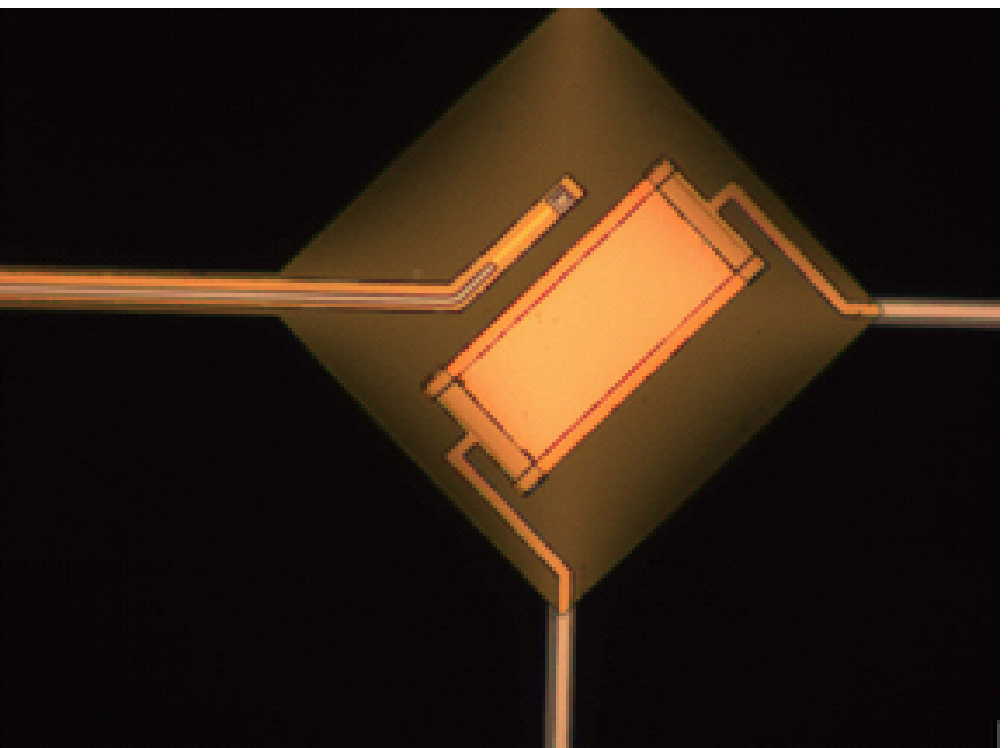}%

\caption{{\it Left:} Layout of prototype Clover detector chip.  The tapered end provides a gradual impedance transition.  {\it Right:} Clover  prototype bolometer silicon nitride island showing TES and microstrip leading to termination resistor.}
\label{fig:chip}
\end{figure}

Power is coupled from the waveguide to the TES planar circuit using an
antipodal finline taper consisting of two superconducting fins of Nb
separated by 400~nm of SiO$_2$ \cite{Yass95,Yass2000} (see
Fig.~\ref{fig:chip}). Before the fins overlap, the thickness of the
SiO$_2$ is much less than that of the silicon substrate and the structure
behaves as a unilateral finline. As the fins overlap, the structure
starts to behave like a parallel-plate waveguide with an effective
width equal to the overlap region. When the width of the overlap
region becomes large enough for fringing effects to be negligible, a
transition to a microstrip mode is performed. The microstrip is then
tapered to the required width.

%-------------

Clover's bolometers are low-stress silicon nitride islands suspended
on four legs (see Fig.~\ref{fig:chip}).  The nitride is 0.5~$\rm\mu m$
thick.  The thermal conductance to the thermal bath is controlled by
the four nitride legs.  The microstrip carrying power from the finline
to the bolometer is terminated by a 20-$\Omega$ Au/Cu resistor which
dissipates the incoming power as heat that the TES can detect.  The
TES films in Clover are Mo/Cu proximity-effect bilayers, with
transitions as sharp as 1--2 mK for high sensitivity.  The operating
temperature of Clover's detectors is chosen to meet the NEP
requirements, which are dominated by phonon noise.

\begin{figure}
\centering
\includegraphics[width=8.0cm]{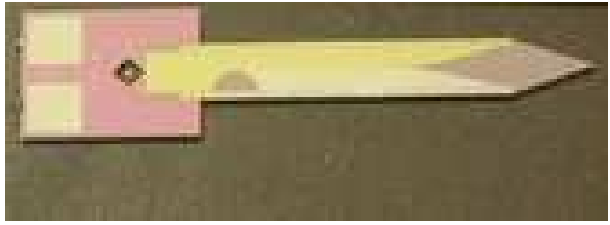}%
\caption{Prototype Clover detector chip.  The chip is about 16~mm long.}
\label{fig:tweezers}
\end{figure}

We have produced prototype detectors for Clover and found that the
TES films are of high quality, making for sensitive detectors,
although more development of the thermal design is needed to achieve
the required power handling.  The measured electrical NEP is close to
the phonon noise level.  Figure~\ref{fig:tweezers} shows a prototype
Clover detector chip.

%-------------

\begin{figure}

\includegraphics[width=6.0cm]{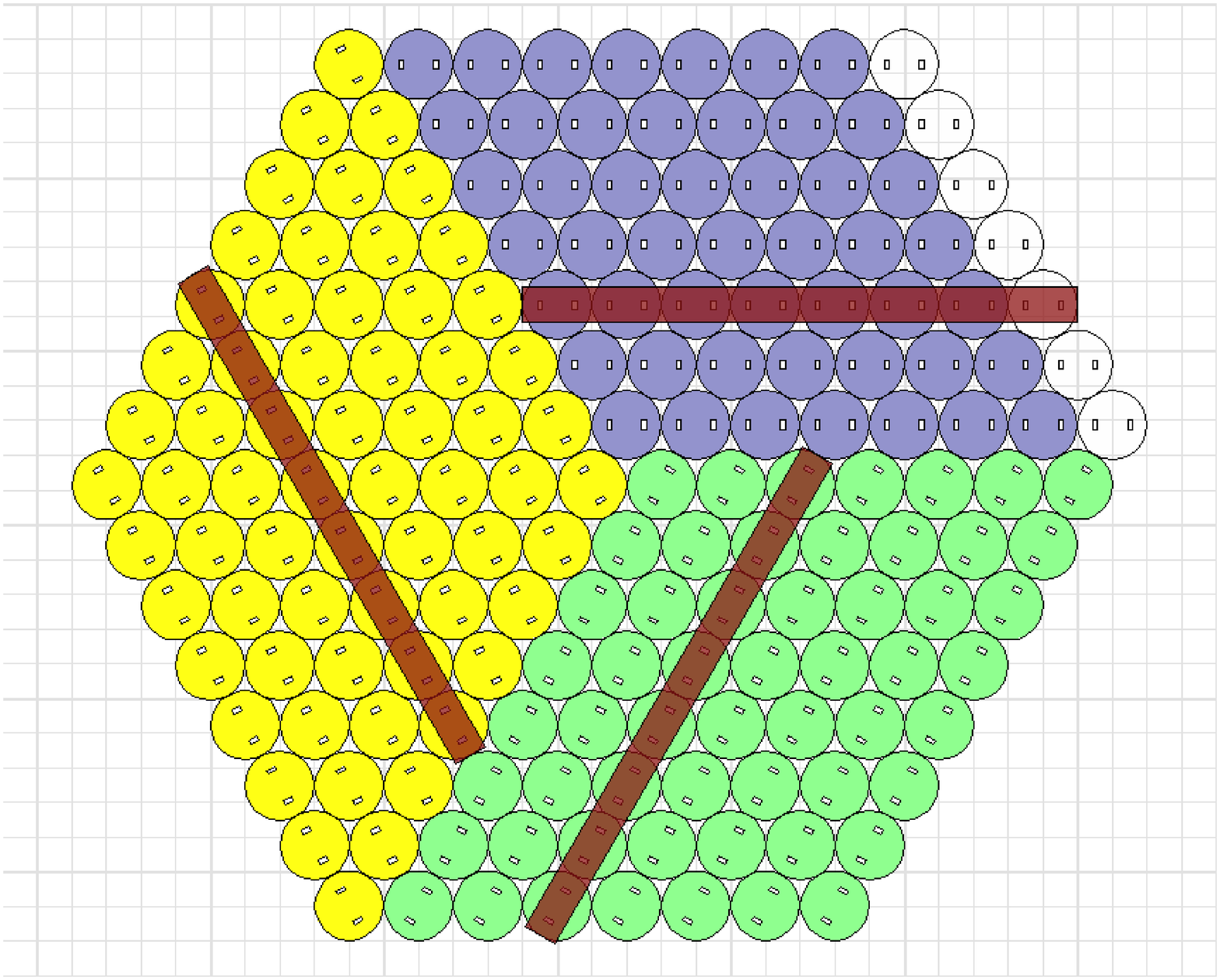}
\hspace{1cm}%
\includegraphics[width=6.0cm]{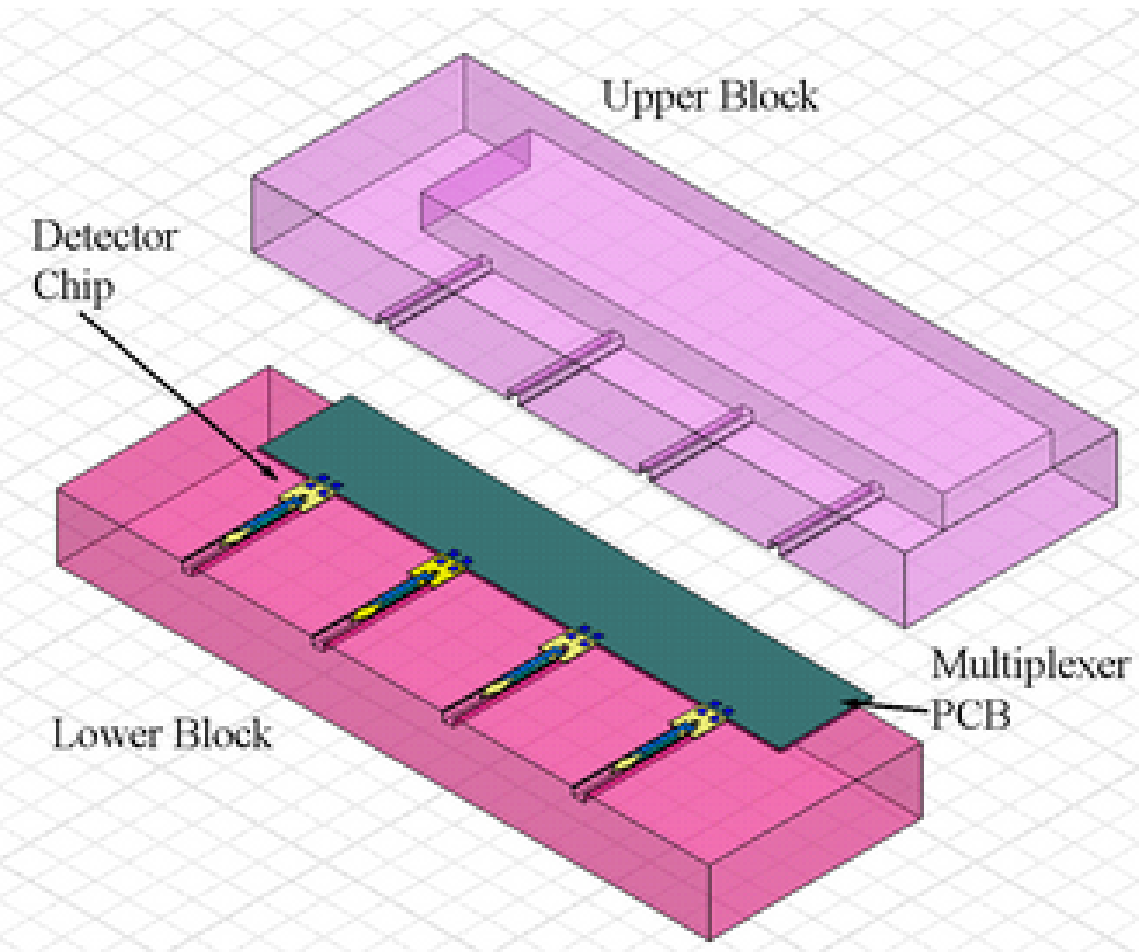}%

\caption{{\it Left:} Layout of 90-GHz focal plane.  The dark
rectangles show the orientation of linear detector blocks.  {\it
Right:} Detector block concept showing how four detectors would be
mounted in a block.  The upper and lower blocks form waveguides in
which the finlines sit.}
\label{fig:hexeight}
\end{figure}

The feedhorns are arranged in a hexagonal array.  However, the
$1\times32$ multiplexer chips we are using lend themselves more
naturally to a planar configuration where we have up to 16 horns in a
row.  As shown in Fig.~\ref{fig:hexeight}, we split the 97-GHz focal
plane into three regions.  The two waveguides corresponding to each
horn are arranged so that they are all parallel within one of these
regions.  This allows us to cover each region with linear detector
blocks stacked on top of each other with an offset to match the
hexagonal horn pitch.  The orientation of one of these detector blocks
is shown by a dark rectangle in each of the three regions.  The 97-GHz
focal plane needs 22 detector blocks to cover it.  The scheme for
covering the 150- and 220-GHz focal planes is similar, except that the
horns are arranged in a hexagon with a side of ten horns.

%-------------
A simplified view of a detector block holding four detectors is shown
in Fig.~\ref{fig:hexeight}.  The detector modules contain 16 or 20
detectors each for compatibility with the hexagonal arrays of horns in
the telescopes' focal planes. The detector block comes in two halves,
upper and lower.  When these are put together they form split-block
waveguides, into which the finlines protrude.

%-------------
\section{Site and Observing Strategy}

Clover is planned to be sited at the Llano de Chajnantor in northern
Chile, at latitude $-23.5$ deg. This site combines relative ease of
access (it can be reached by public roads) with superb millimetre and
sub-millimetre observing conditions, and is the site of many planned
or existing instruments (e.g.\ CBI, QUIET, ALMA, APEX, ACT).

\begin{figure}
  \centering
  \includegraphics[height=0.4\textwidth]{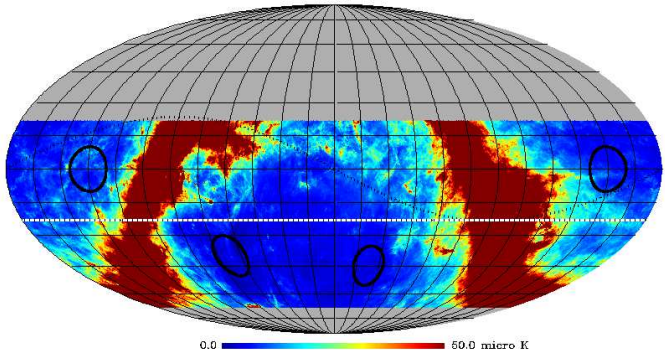}
  \caption{\label{fig:field} The predicted total
    intensity of synchrotron and dust emission at 97GHz with $10^\circ$
    fields selected for observations from Chajnantor overlaid.}
\end{figure}

Clover will likely observe observe four patches of sky, each of radius
$\sim 10^\circ$, distributed as equally as possible in right ascension
in order to provide year-round useful observing. The exact fields will
be chosen to be in areas identified to have low total-intensity
emission and predicted to have low levels of diffuse, polarized
galactic foregrounds (synchrotron and thermal dust emission).  While
not much is known about low-level polarized foregrounds, models do
exist (e.g. \cite{giardino}, \cite{SFD}), which we extrapolate to
obtain estimates in our observing bands. The sky patches must also be
observable at high elevation for long periods, but not go overhead,
since constant-elevation scanning is not possible at the zenith. From
Chajnantor there are two general areas that fulfil this requirement,
one in each Galactic hemisphere, at RA = 21h -- 05h, Dec $= -45$ and
RA = 08h -- 14h DEC $= 0$.  Fig.~\ref{fig:field} shows the predicted
total intensity of synchrotron and dust emission at 97 GHz and the
positions of four suitable $10^\circ$ fields visible from Chajnantor.

\begin{figure}[t!]
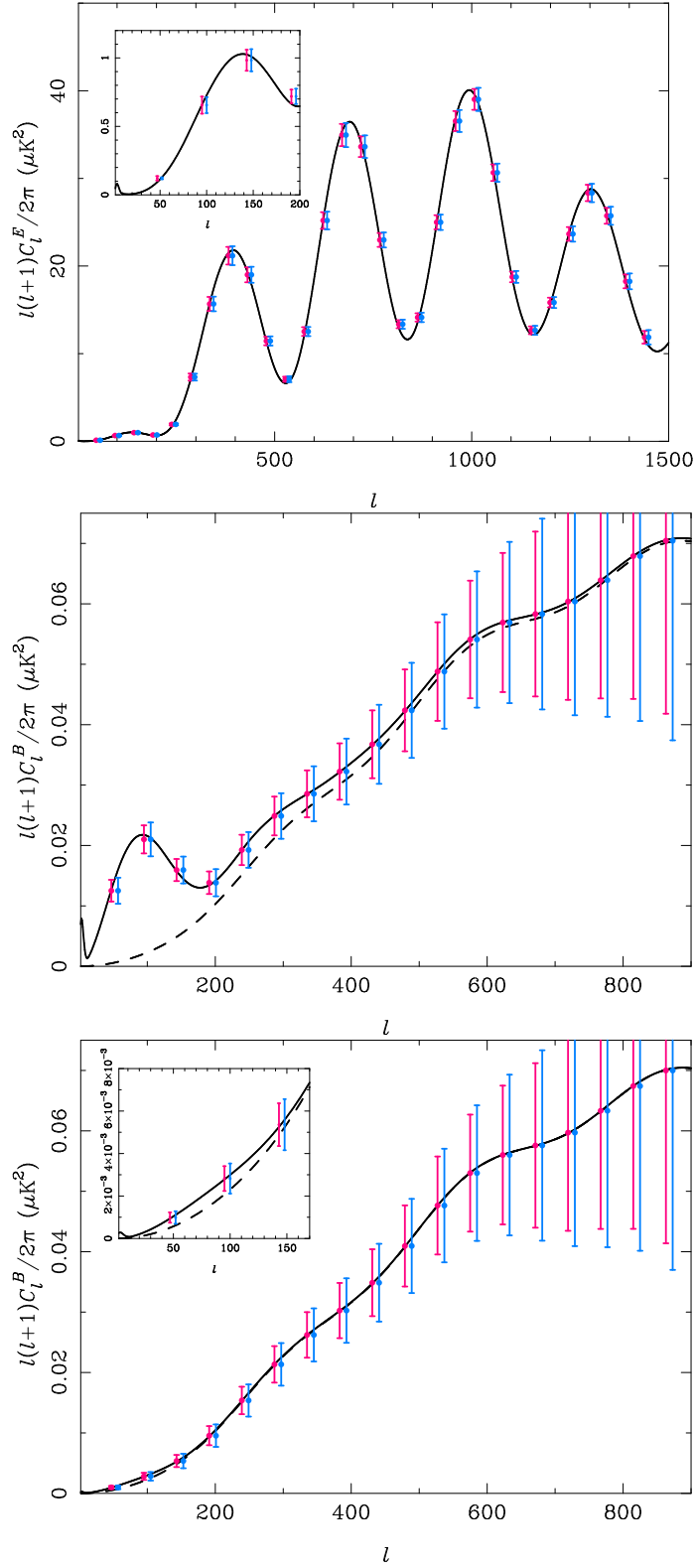

\begin{center}
\includegraphics[height=0.4\textheight,angle=-90]{Clover_errors_Chile2yr_4x8p9deg_rp01_E}
\includegraphics[height=0.4\textheight,angle=-90]{Clover_errors_Chile2yr_4x8p9deg_rp28_B}
\includegraphics[height=0.4\textheight,angle=-90]{Clover_errors_Chile2yr_4x8p9deg_rp01_B}
\end{center}
\caption{\label{fig:atacama} Power spectrum error forecasts after two years
of operation at Chajnantor for $E$ modes (top) and $B$ modes with
$r=0.28$ (middle)
and $r=0.01$ (bottom). The blue error bars are the results of the exact
Fisher matrix computation; the magenta errors are approximations that
ignore the effects of $E$-$B$ mixing due to the finite survey geometry.
Note that on large scales, where the mixing is most important, the approximate
error bars overestimate the errors on $C_l^E$. For $C_l^B$, ignoring
the mixing leads to overly-optimistic errors on large scales. The dashed
line in the $C_l^B$ plots is the contribution from gravitational lensing.
} 
\end{figure}

Our adopted scanning strategy is based on scanning at constant
elevation, which eliminates large baselines on each scan due to
changing airmass. It is important that each field can be observed with
scans at a wide range of position angles. This is necessary for the
mapping strategy, which uses the redundancy introduced by the
cross-coverage to fit out and remove low-order components along each
scan due to residual atmospheric fluctuations or receiver
instabilities. A range of crossing angles will also be important for
averaging down any polarization systematics.

%- ----
\section{Power spectrum and parameter forecasts}

The errors on the power spectra and the tensor-to-scalar ratio $r$
have been forecasted using a purpose-written Fisher-matrix code. This
properly accounts for the effect of the survey geometry on the power
spectra errors and their covariance. In particular, it includes
information loss in the $C_l^B$ errors due to $E$-$B$ mixing due to
the finite sky coverage and the small effect of (anti-)correlations
between neighbouring bandpower estimates. We assume that there are no
significant pixel-pixel correlations, which should be the case for
Clover maps after de-striping, and we have not attempted to account
properly for the change in noise properties in the maps due to
foreground removal, nor do we take account of any foreground
residuals. Since the polarized foregrounds at Clover frequencies are
still highly uncertain, our noise figures are based solely on the
middle, 150-GHz channel which crudely (but consistently) accounts for
removal of synchrotron emission with the lowest frequency and thermal
dust emission with the highest.  We include the $B$-mode lensing
signal as a Gaussian noise term which is approximately correct on the
large scales where the gravitational-wave signal resides and where the
Clover thermal noise level is comparable to the lensing signal. We
have not attempted to clean out the lens-induced $B$-modes although
this should be possible to some extent with Clover data and would
reduce the errors a little on $C_l^B$.

Our forecasts for the power spectrum errors after two years of
operation are presented in Fig.~\ref{fig:atacama}. The total survey
area is $1000\,\mathrm{deg}^2$. This is close to optimal in terms of
balancing thermal variance against the sample variance of the
lens-induced $B$ modes. We have assumed year-round, night-time
observing with an average observing efficiency (taking into account
bad weather, down time and calibration) of about 40 percent. In each
case we show the one-sigma errors on bandpowers $\mathcal{C}_l
\equiv l(l+1)C_l/(2\pi)$, after marginalising over neighbouring bands.
We adopt two values for the tensor-to-scalar ratio: $r=0.28$, the
current 95\% limit from WMAP3 temperature and $E$-mode polarization
data, and $r=0.01$ which is the design target for Clover.

Propagating the band-power errors in $C_l^B$, after marginalising over
$C_l^E$, to errors in the tensor-to-scalar ratio gives a one-sigma
error on $r$ in the null hypothesis of 0.004.  Marginalising over
$E$-modes produces only a very small degradation compared to when we
include them in the analysis showing that the constraint on $r$ is
indeed driven by the $B$-modes. We only allow $r$ to vary which
neglects the small uncertainty in the lens-induced $B$-mode power.

We end by noting that the angular resolution of Clover (8 arcmin) also 
benefits the secondary science that we can do with Clover. Examples include
lensing reconstruction from the polarization maps and hence constraints
on dark-sector parameters (e.g. neutrino masses and dark energy properties),
and searches for the imprint of cosmic (super)strings or primordial
magnetic fields in the $B$-mode power spectrum on intermediate scales.
In addition, the improved angular resolution of the Clover survey will
improve the Galactic science (e.g. magnetic field morphology and dust grain
properties) returned from Clover.

\section*{Acknowledgements}
Clover is funded by the Particle Physics and Astronomy Research
Council. ACT is supported by a PPARC fellowship. The work described
here is the result of many peoples' effort within the Clover
collaboration. Particular thanks to Damian Audley and Anthony
Challinor for help with these proceedings.

\end{document}